\documentclass[10pt,conference]{IEEEtran}
\IEEEoverridecommandlockouts
\usepackage{hyperref}
\hypersetup{
    colorlinks=true,
    linkcolor=blue,
    urlcolor=cyan,
    citecolor=blue,
}

\usepackage{balance}
\usepackage{amsmath}
\usepackage{amssymb}
\usepackage{amsfonts}
\usepackage{algorithmic}
\usepackage{graphicx}
\usepackage{textcomp}
\usepackage{xcolor}
\usepackage{algorithmic}
\usepackage{graphicx}
\usepackage{textcomp}
\usepackage{xcolor}
\usepackage{verbatim}
\usepackage{listings}
\usepackage{multirow}
\usepackage{enumitem}
\usepackage{framed}
\usepackage{colortbl}
\usepackage{soul}
\usepackage{booktabs} 
\usepackage{tcolorbox}

\usepackage[numbers,sort&compress]{natbib} % Use natbib for IEEE-style numbered citations

\definecolor{lightgray}{gray}{0.95}
\definecolor{darkblue}{rgb}{0,0,0.5}
\definecolor{lightred}{rgb}{1, 0.8, 0.8}
\definecolor{lightyellow}{rgb}{1, 1, 0.8}
\definecolor{lightgreen}{rgb}{0.6, 1, 0.6}
\definecolor{mint}{rgb}{0.74, 0.99, 0.79}
\definecolor{forestgreen}{rgb}{0.13, 0.55, 0.13}
\definecolor{goldenrod}{RGB}{255, 185, 65}

\def\BibTeX{{\rm B\kern-.05em{\sc i\kern-.025em b}\kern-.08em
    T\kern-.1667em\lower.7ex\hbox{E}\kern-.125emX}}
\begin{document}

\title{Cracks in The Stack: Hidden Vulnerabilities and Licensing Risks in LLM Pre-Training Datasets
 \thanks{Replication package available at: \url{https://zenodo.org/records/14175945}}
}

\author{
\IEEEauthorblockN{Mahmoud Jahanshahi, Audris Mockus}
\IEEEauthorblockA{\textit{Department of Electrical Engineering and Computer Science} \\
\textit{University of Tennessee, Knoxville, USA} \\
\{mjahansh, audris\}@utk.edu}
}

\maketitle

\begin{abstract}

A critical part of creating code suggestion systems is the 
pre-training of Large Language Models (LLMs) on vast amounts 
of source code and natural language text, often of questionable
origin, quality, or compliance. 
This may contribute to the presence of bugs and vulnerabilities
in code generated by LLMs. 
While efforts to identify bugs at or after code generation 
exist, it is preferable to pre-train or fine-tune LLMs on 
curated, high-quality, and compliant datasets. 
The need for vast amounts of training data necessitates 
that such curation be automated, minimizing human intervention.

We propose an automated source code autocuration technique
that leverages the complete version history of open-source
software (OSS) projects to improve the quality of training data.
The proposed approach leverages the version history of all 
OSS projects to: (1) identify training data samples that 
have ever been modified, (2) detect samples that have undergone 
changes in at least one OSS project, and (3) pinpoint a 
subset of samples that include fixes for bugs or vulnerabilities.
We evaluate this method using ``the Stack'' v2 dataset, 
comprising almost 600M code samples, and find that 17\% of
the code versions in the dataset have newer versions, with 17\%
of those representing bug fixes, including 2.36\% addressing
known CVEs. 
The clean, deduplicated version of Stack v2 still includes 
blobs vulnerable to 6,947 known CVEs. 
Furthermore, 58\% of the blobs in the dataset were never 
modified after creation, suggesting they likely represent
software with minimal or no use. Misidentified blob origins
present an additional challenge, as they lead to the 
inclusion of non-permissively licensed code, raising serious 
compliance concerns.

By deploying these fixes and addressing compliance issues,
the training of new models can avoid perpetuating buggy code 
patterns or license violations. We expect our results to 
inspire process improvements for automated data curation, 
a critical component of AI engineering, with the potential
to significantly enhance the quality and reliability of 
outputs generated by AI tools.
\end{abstract}

\begin{IEEEkeywords}
Large Language Models (LLMs), The Stack v2 Dataset, Open Source Software (OSS), LLMs for Code (LLM4Code), Software Supply Chains, World of Code (WoC), Security Vulnerability, Open Source Licensing
\end{IEEEkeywords}

\section{Introduction}\label{intro}

Large Language Models (LLMs) are already employed by popular tools
such as GitHub Copilot and have a significant impact on how people
interact with computing resources.
LLM code-generation tools appear
to increase productivity~\cite{ziegler2022productivity}, are easy to
access with little or no cost on popular coding platforms, and
generated code is rapidly spreading~(``GitHub Copilot is behind an
average of 46\% of a developers'
code''~\cite{zhao2023github}). 
Quality control of this code, however, is severely lacking in the
LLM-based Software Supply Chain (SSC).
LLMs are trained on vast
amounts of source code and natural language text that are of
questionable origin and quality.
The output generated by LLMs,
therefore, often contains bugs, vulnerabilities, or license
violations that are copied or reused to train other LLM models, thus
propagating the problem.
~\citet{hubinger2024sleeper}
showed that LLMs can introduce vulnerabilities and this behavior is
extremely difficult to change via
fine-tuning.
It is reasonable to assume
that at least part of that buggy output may be attributed to the
buggy files used to train LLMs.
While existing approaches use AI
to detect the most common insecure coding patterns~\cite{zhao2023github}, but
many vulnerabilities do not fit such simple patterns.
It is widely accepted that the size and quality of training corpus
are essential for good performance of the models, yet common
curation techniques, such as number of stars or forks, appear
ineffective~\cite{allal2023santacoder}.
Independent of the intended coding tasks, a large body of training
data is necessary for LLMs to be effective.
As poor quality training
data can reduce the quality of LLM-based tools, improving the state
of art in source code training data curation is an important task
that would impact all downstream efforts.
It is worth noting that
source code is often included in training data for natural language
models as well.
For example, the natural language collection
in~\cite{laurenccon2022bigscience} has hundreds of gigabytes of source
code and collection described in~\cite{gao2020pile} nearly 100GB.

Previous work found instances of vulnerable or
license-violating code in open source training datasets.
This shows
that by taking information from version control systems, it is
feasible to identify vulnerable, buggy, or license-violating code
and replace it with fixed versions~\cite{reid2022extent,reid2023applying}.

In summary,
it is essential to exclude problematic code from LLM training
datasets, or, at least, to flag it as high risk.

The goals of this work is to investigate the quality of the source
codes that are used to train LLMs and to develop automated
approaches to improve it. 
Specifically, we propose a simple and
effective way to identify (and fix) several types of problematic
source code that is used to train LLMs.

In a nutshell, we leverage the fact that a file’s content may 
undergo numerous changes over its lifetime, with some of these 
changes being bug fixes. 
By identifying cases where a file in the training data has 
been modified and updated, we can recommend these newer 
versions as replacements for older versions in the training dataset.
In order for this approach to work, we have to go
across repository boundaries and consider versions (and their
history) in all public repositories, i.e., Universal Version History
(UVH)~\cite{reid2023applying}.
World of Code (WoC) research
infrastructure~\cite{ma2019world,ma2021world} provides capabilities to
accomplish such an arduous task as described in
Section~\ref{s:method}.

Our primary contributions are: 1) an approach to identify
potentially vulnerable, buggy, or not heavily used source code in
public LLM training datasets; 
2) an approach to identify potential license violations in these datasets; and
3) evaluation of the approach on the largest
public curated code LLM training dataset the Stack
v2~\cite{lozhkov2024starcoder}. 
We also articulate how code LLM's
represent a novel type of software supply chains and suggest that
never-modified code may indicate its low use and untested quality
and that should be taken into account when constructing training
datasets.

In the remainder of the paper Section~\ref{s:background} discusses
curated training datasets used for evaluation, relevant key concepts
of software supply chains, how LLM-generated coder represents a
novel type of software supply chain, and key features of WoC used in
this study.
Section~\ref{s:method} describes our approach in detail.
Section~\ref{s:results} presents and discusses
our findings.

\section{Background}\label{s:background}

\subsection{Types of Software Source Code Supply Chains}~\label{s:ssc}
Software supply chain concept is helpful for assessing risks, as in
traditional supply chains. However, software supply chains have
substantially different nature from traditional supply chains.
In particular, three types of software source code\footnote{We
  explicitly exclude various ways binary software is delivered as,
  for example in Solar Winds hack.} supply chains have been
previously identified~\cite{mockus2019insights}. 
The most common, or Type I SSC
is represented by code (runtime) dependencies. For example, an
import statement in Java or include statement in C programming
languages. The two primary risks for downstream projects in this 
scenario are: insufficient upstream maintenance, where bugs and 
vulnerabilities remain unresolved, and overly aggressive maintenance, 
where upstream changes disrupt downstream code~\cite{xavier2017historical}.  

Type II SSC involves copied code, a common practice in open-source 
software where code is shared publicly~\cite{jahanshahi2024beyond}, 
allowing anyone to copy or fork it (within licensing requirements). 
While breaking changes are no longer a risk in Type II SSC, the 
absence of upstream maintenance becomes inevitable, as the code 
is now maintained within the destination project.

Type III SSC involves knowledge transfer where developers learn
procedures techniques and tools by working in one project and then
apply some of what they learned elsewhere. While learning, in
general, is a good thing, some quality practices or API usage may
introduce bugs or vulnerabilities that, if adopted by developers, are then spread
by these developers to other projects.

The current state of the industry in source code SSCs is to
capture dependencies based on package managers (Type I SSCs) and to rely on the
``official'' directories such as NVD and package managers to
identify the security and licensing attributes. As was shown
in~\cite{reid2022extent,reid2023applying}, rampant code copying 
enabled and encouraged
by OSS results in massive orphan vulnerabilities and licensing
violations that cannot be detected by existing approaches.

\subsection{The Promise and Challenges of Large Code Datasets}

Large-scale code datasets are invaluable for advancing AI-driven code solutions, such as automated code generation, bug detection, and refactoring.
These datasets provide extensive repositories of programming languages, styles, and structures, enabling large language models (LLMs) to learn complex coding patterns and generalize across diverse coding tasks.
By leveraging such data, AI models significantly improve in generating, completing, and correcting code, which supports developers in accelerating the software development cycle and reducing costs~\cite{allamanis2018survey,lozhkov2024starcoder}.

However, maintaining the quality and integrity of these large datasets poses several challenges, often underexplored in research.
Duplication, for instance, can lead to redundancy, creating biases and reducing model diversity.
Version control is another critical challenge, as datasets sourced from dynamic platforms like GitHub may frequently change;
without careful version tracking, models risk learning outdated or deprecated practices.
Provenance tracking is essential for maintaining the contextual relevance and reliability of data, allowing users to trace the origins and evolution of code snippets.
Additionally, licensing complexities arise, as open-source code often comes with a range of permissive and restrictive licenses.
Properly handling these licensing issues is crucial to ensuring lawful usage, especially in commercial settings~\cite{gunasekar2023textbooks}.

LLMs introduce a novel type (Type IV) of Software Supply Chains
that manifest by relationships
between the LLM-generated code and the code used to train the LLM
models. LLMSSCs, similar to Type II SSCs, are conceptually copying
the code (including its bugs) in the training data but in a way that
obfuscates the origin. The full scope of risks posed by Type II
copy-based SSCs has yet to be studied in depth.

\subsection{The Stack v2 Dataset}

To evaluate our approach we use a large open source dataset
intentionally curated for training code LLMs: the Stack
v2~\cite{lozhkov2024starcoder}. ``The Stack v2 contains over 3B
files in 600+ programming and markup languages. The dataset was
created as part of the BigCode Project , an open scientific
collaboration working on the responsible development of Large
Language Models for Code (Code LLMs). The Stack serves as a
pre-training dataset for Code LLMs, i.e., code-generating AI systems
which enable the synthesis of programs from natural language
descriptions as well as other from code snippets.''  

This dataset is widely adopted in AI and software development 
due to its extensive multi-language coverage and permissive licensing, 
enabling use in both academic and commercial contexts.
The Stack (v2) fosters open collaboration, supporting model 
training across diverse coding ecosystems and advancing tools 
for software automation and analysis~\cite{gunasekar2023textbooks}.

In addition to
the full dataset, the Stack v2 has several deduplicated versions.
the-stack-v2-dedup is near-deduplicated, the-stack-v2-train-full-ids
is based on the the-stack-v2-dedup dataset but further filtered with
heuristics and spanning 600+ programming languages. Finally,
the-stack-v2-train-smol-ids is based on the the-stack-v2-dedup
dataset but further filtered with heuristics and spanning 17
programming languages. We evaluate our fixing approach on the \textbf{full}
and \textbf{smol} (maximally deduplicated) datasets\footnote{
For more details on the dataset and the deduplication process, refer to the 
official Stack v2 documentation: \url{https://huggingface.co/datasets/bigcode/the-stack-v2}
}.

\subsection{Motivation for This Study}

Evaluating large code datasets is essential to address the intricacies of version security and licensing, which collectively impact the reliability and ethical compliance of large language models (LLMs) for code.

\subsubsection{Security Vulnerabilities and Bugs} The security implications of large datasets are significant, especially in the context of outdated or vulnerable code.
If models are trained on datasets containing undetected security flaws, these vulnerabilities may persist in model outputs, increasing the risk of insecure code suggestions.
This issue is particularly concerning for code used in sensitive applications, where even minor security oversights can lead to substantial risks and exploitation potential.
Security-focused dataset evaluation is therefore vital to prevent models from inadvertently embedding insecure practices into their code outputs~\cite{pearce2022asleep}.

Given the large-scale, open-source nature of The Stack v2 dataset, it is likely to contain instances of vulnerable and buggy code.
This hypothesis (\textbf{H1}) is based on the prevalence of ``orphan vulnerabilities'' in open-source projects, as described by~\citet{reid2022extent}, where vulnerabilities in copied code persist even after they are patched in the original source.
In large datasets aggregated from numerous repositories, code reuse without consistent patching introduces security risks, as outdated or unpatched code versions may proliferate across projects, spreading known vulnerabilities further.

\begin{itemize}
    \item \textbf{Hypothesis 1 (H1)}: The Stack v2 dataset is likely to contain instances of vulnerable and buggy code.
\end{itemize}

\subsubsection{Legal Considerations} Maintaining licensing integrity is fundamental for the lawful and ethical deployment of code-based AI.
The provenance and licensing of code samples in these datasets must be meticulously tracked to prevent legal risks associated with licensing misrepresentation or inaccurate attributions. 
Open-source projects often involve significant code reuse, which can lead to fragmented metadata or altered licensing information as code is copied across projects.
Proper licensing ensures that the models’ outputs respect open-source constraints, which is crucial for both research and commercial applications.
Without rigorous checks, models might generate code based on improperly licensed data, exposing end-users to compliance issues and potential litigation.
Ensuring that datasets uphold licensing integrity not only fosters ethical AI but also protects users from unforeseen legal complications~\cite{gunasekar2023textbooks}.

Due to the prevalence of ``copy-based reuse'' in open-source development, as explored by~\citet{jahanshahi2024beyond}, we hypothesize (\textbf{H2}) that The Stack v2 dataset contains instances of misidentified code origins. While the dataset has metadata identifying the project from where each source code file was obtained, that file may have been copied from another project that has a different or even incompatible license.  
This form of reuse, where source code is directly copied into new projects, often results in fragments with altered or lost metadata, which complicates the ability to accurately trace their provenance.
This lack of provenance tracking can lead to legal and ethical issues in AI applications for code.
Without accurate metadata, models may inadvertently generate code with improper licensing, exposing users to potential compliance issues.
Misidentification of code origins in datasets like The Stack v2 is particularly risky for industry applications, as it challenges the trustworthiness and lawful deployment of LLM4Code models in commercial environments.

\begin{itemize}
    \item \textbf{Hypothesis 2 (H2)}: The Stack v2 dataset is likely to contain instances of misidentified code origins that are prone to license violation.
\end{itemize}

\subsection{Contributions}

The primary contributions of this paper focus on addressing  data quality and compliance concerns within The Stack v2 dataset.
The paper aims to enhance the understanding and reliability of large code datasets by providing the following key contributions.

\subsubsection{Assessment of Security and Reliability} 

We introduce a novel methodology for identifying source code that may be potentially vulnerable, contain bugs, or exhibit minimal usage in real-world applications. 
Our approach uniquely incorporates version control history to track and analyze the evolution of source code, focusing on identifying newer versions of files that indicate updates, bug fixes, or refinements over time. By examining commit histories and versioning patterns, we can detect files that have undergone improvements or corrections, flagging older versions as potentially vulnerable or buggy. This historical perspective provides insight into code stability and usage trends, allowing us to differentiate actively maintained, reliable code from outdated, less robust sections. 

\subsubsection{Analysis of Code Provenance and Licensing Accuracy} 

We conduct a detailed examination of code provenance to evaluate licensing accuracy and the origins of code snippets within the dataset.
By tracking the source and licensing status of code entries, we provide a comprehensive assessment of compliance with open-source licensing requirements.
This contribution is particularly important for models deployed in industry, where legal and ethical use of data must be assured.

\subsubsection{Evaluation on Large-Scale Code Dataset}

To validate the effectiveness of our approach, we perform a comprehensive evaluation on the largest publicly curated code LLM training dataset, Stack v2. This dataset serves as an ideal benchmark due to its scale and diversity. By applying our methodology to Stack v2, we can assess the robustness of our techniques in identifying potentially vulnerable or outdated code segments, accurately tracking version histories, and verifying licensing compliance across a large and varied dataset. This evaluation establishes the applicability and scalability of our contributions to real-world, large-scale code datasets, reinforcing the value of our work in supporting the development of secure, high-integrity LLM training corpora.

\section{Methodology}\label{s:method}

To address big data-related aspects of the proposed work, we 
leverage WoC research infrastructure~\cite{ma2019world,ma2021world} for
open source version control data. This data includes a vast majority
of public open source projects and provides access to petabytes of data
that includes versions of source code, information on time,
authorship, and exact changes made to the source code over the
entire activity history of most participants in OSS.

\subsection{Key Concepts}
The proposed method for identifying issues in training data leverages 
unique capabilities of WoC.  
In particular, WoC's ability to cross-reference and track the 
history of code versions across nearly all public
repositories, along with its curated data that addresses complex 
challenges like repository deforking~\cite{mockus2020complete} 
and author ID aliasing~\cite{fry2020dataset}, makes this approach 
feasible.

We use a simple example to demonstrate the tracing and cross-referencing 
capabilities of WoC. Suppose we take a single sample $b$ (version or, in 
git terms, blob) of source code 
from any training (or test) data. We can calculate git SHA-1~\footnote{
Git SHA-1 is simply a SHA-1 calculated on the string (representing the content) 
with prepended string ``blob SIZE\textbackslash0'' where SIZE 
is the length of the content.}
for this sample. All further calculations use 
git SHA-1 and do not require the content.

For a blob $b$ to materialize in a version control repository, it has to be created by 
a commit $c$. Git commits include the time of the commit, 
commit message, SHA-1 of the parent commit(s) and 
SHA-1 of the tree (folder). WoC, by comparing the trees\footnote{WoC contains 
over 20B blobs.} of the commit and its parent(s) determines all 
the modifications to the project done by the commit. Specifically, in 
case any of the project's files are modified, it extracts 
the tuple $(b_o, b_m)$ representing the old and the new version of the file. 
These pairs are associated with the commit and its other 
attributes, like time, author and commit message. 

Suppose there is a commit, $c_t(b_o, b_m)$, which 
addresses a vulnerability $v$ in project $P$. This commit, $c$, 
modifies a file $f$ at time $t$, where the original version of the 
file is represented by the blob $b_o$ and the modified version 
by $b_m$. WoC’s cross-referencing allows us to identify all 
repositories containing $b_o$ or $b_m$, all relevant commits, 
their parent and child commits, and the authors and projects 
associated with these commits.

Typically, we need a repository and a commit to identify what files
were changed, their content before and after the change, as well
as the parent commit. By collecting and cross-referencing nearly all
open source data, WoC allows us not only to go forward in version
history (see child commits), but also to identify all commits that
either created or modified a particular version of the file. To
identify problems with the LLM training data, we will first match it to
blobs or commits in WoC. Both the Stack and the Stack v2 contain
versions of the files (blobs) and their git SHA-1 digests. We,
therefore, just need the list of SHA-1 digests to match them to
blobs in WoC. We further assume that if there exists at least one commit that 
modifies $b_o$, and its commit log message contains 
keywords (described below) indicating that 
it is a fix, then that blob is buggy. Similarly, if the 
commit indicates that it fixes a vulnerability, we assume 
that modified blob contains vulnerability.

\subsection{Identifying Potential Noncompliance}\label{s:noncompliance}

The Stack dataset provides information on repositories and their 
identified licenses for all blobs. Since code reuse through copying 
is common among developers~\cite{jahanshahi2024beyond}, accurately 
tracing the originating projects for each blob can be challenging. 
WoC addresses this by offering a map~\cite{jahanshahi2024dataset} 
that, for blobs found in multiple projects, sorts them by the commit 
time of each blob’s creation, allowing us to identify its first 
occurrence and the repository where it was initially committed. 
By comparing this origin information from WoC with the data in 
the Stack, we can verify whether the originating repository of
each blob has been accurately identified.

If the origin identified by WoC does not match the origin listed
in the Stack data, we then analyze the licenses associated with 
both the WoC-identified originating repository and those detected 
by the Stack. Using WoC's license map~\cite{jahanshahi2024license}, 
we compare this information with the Stack’s license data to 
identify potential instances of license noncompliance.

\subsection{Sampling}

We used a $\frac{1}{128}$th sample for certain quantitative analyses to 
balance computational feasibility with representativeness. 
The sampling was based on SHA-1 hashes of the blobs and commits, 
which ensures that the selection process is effectively random. 
This approach maintains statistical robustness while significantly 
reducing the computational overhead of processing the entire dataset.

\section{Results and Discussions}\label{s:results}

\subsection{Hidden Vulnerabilities}

As described in Section~\ref{s:method}, we first extract git SHA-1
for all blobs in the Stack v2 (\textit{full}) and the-stack-v2-train-smol-ids
(\textit{smol}) datasets. The former has 582,933,549 and the latter
has 87,175,702 unique
blobs. The total number of blobs in WoC version V3 (extracted at
about the same time as the Stack v2) has over 26B blobs, or almost
45 times more blobs than the full version and 300 times more than the
small deduplicated version.

Starting from these two lists of blobs\footnote{The second list had only
26\% overlap with the first list instead of being a strict subset of
the first.}
we first obtained two maps to commits: the first map links
blobs to commits creating the blob (including the previous version of the
file), while the second map links to commits that modified the file,
thereby creating a new blob, as described in the previous section.
Not all blobs could be mapped to commits, as a small fraction did 
not appear in either map. This could be due to certain code versions 
being created without a publicly accessible version history or missing 
corresponding commits or trees in WoC.

Table~\ref{t:counts} summarizes the blob counts for two evaluation 
datasets, based on a $\frac{1}{128}$th random sample determined by 
the SHA-1 hash of each blob. These counts can be extrapolated to 
the full dataset by multiplying by 128. 

From Table~\ref{t:counts}, we observe that approximately $2.5\%$ 
of the blobs could not be linked to any commits. Among the 
remaining blobs, $62\%$ and $55\%$ represent files that were created 
without preceding blobs, i.e., they are the initial versions. Of these, 
only $5.5\%$ and $4.6\%$ had a newer version, meaning the majority were 
created but never modified. Since the first version of frequently 
executed source code is rarely error-free, this lack of updates 
suggests the code was likely not used in practice, raising 
concerns about its overall quality.

\begin{table}[t]
  \begin{center}
    \caption{Counts in the blob sample}
    \label{t:counts}
    \resizebox{0.99\linewidth}{!}{
    \begin{tabular}{ll|rrrr}
    \toprule
    & & \multicolumn{2}{c}{\textbf{full}} & \multicolumn{2}{c}{\textbf{smol}} \\
    & & count & \% (row) & count & \% (row)\\
    \midrule
    1 & \textbf{Total} & \multicolumn{2}{c}{\textbf{4,553,119}} & \multicolumn{2}{c}{\textbf{680,917}} \\
    2 & Missing & 115,239 & 2.53 (1) & 16,533 & 2.42 (1) \\
    \midrule
    3 & \textbf{Have an old version} & 1,622,641 & 35.63 (1) & 287,412 & 42.20 (1) \\
    \multicolumn{6}{c}{\dotfill} \\
    4 & \textbf{First version} & 2,813,171 & 61.78 (1) & 376,719 & 55.32 (1) \\
%    5 & Have a new version & 154,366 & 5.48 (4) & 17,339 & 4.60 (4) \\
    5 & No new version & 2,658,805 & 94.51 (4) & 359,380 & 95.39 (4) \\
    \midrule
    6 & \textbf{Have a new version} & 788,059 & \textbf{17.30} (1) & 69,346 & \textbf{10.18} (1) \\
    7 & Found new versions & 1,462,363 & - & 111,453 & - \\
    \bottomrule
    \end{tabular}}
  \end{center}
\end{table}

Furthermore $17.3\%$ and $10.2\%$ of the blobs have a subsequent version(s).
These versions are likely fixing existing bugs, vulnerabilities, make code compatible with newer versions of libraries, or
add new functionality. Since the next version of the code is known,
it would make sense to replace the versions of the training data
with updated versions. 

We further analyze the blobs that have been updated. Using the methodology
described in~\cite{mockus2000identifying}, we identify likely bug fixes 
by searching for terms \textit{fix}, \textit{bug}, \textit{issue},
\textit{patch}, \textit{error}, \textit{resolve}, \textit{correct}, \textit{problem}, 
and their common variations, as well as \textit{cve} in the commit messages\footnote{
\texttt{grep -iwE 'fix|fixes|fixing|bug|bugs|issue|issues|
patch|patches|error|errors|resolve|resolved|resolving|
correct|corrects|corrected|correcting|problem|
problems|debug|debugs|debugged|debugging|cve'}}.

The results are shown in Table~\ref{t:bugs}.
It summarizes the counts for two evaluation 
datasets, based on a $\frac{1}{128}$th random sample determined by 
the SHA-1 hash of each commit that introduces a new version for a blob in the Stack dataset. These counts similarly can be extrapolated to the full dataset by multiplying by 128. 

\begin{table}[t]
  \begin{center}
    \caption{Counts in the new version commit sample}
    \label{t:bugs}
    \resizebox{0.99\linewidth}{!}{
    \begin{tabular}{ll|rrrr}
    \toprule
    & & \multicolumn{2}{c}{\textbf{full}} & \multicolumn{2}{c}{\textbf{smol}} \\
    & & count & \% (row) & count & \% (row)\\
    \midrule
    1 & Commits & \multicolumn{2}{c}{835,699} & \multicolumn{2}{c}{104,782} \\
    2 & Blobs & \multicolumn{2}{c}{5,068,635} & \multicolumn{2}{c}{279,652} \\
    3 & New versions &  \multicolumn{2}{c}{5,657,384} & \multicolumn{2}{c}{307,362} \\
    \midrule
    4 & Fix commits & 137,091 & 16.40 (1) & 13,628 & 13.00 (1) \\
    5 & Fix blobs & 877,811 & \textbf{17.31} (2) & 40,168 & \textbf{14.36} (2) \\
    6 & Fix new versions & 935,587 & 16.53 (3) & 41,222 & 13.41 (3) \\
    \midrule
    7 & CVE commits & 845 & 0.61 (4) & 83 & 0.60 (4) \\
    8 & CVE blobs & 20,765 & \textbf{2.36} (5) & 756 & \textbf{1.88} (5) \\
    9 & CVE new versions & 20,561 & 2.19 (6) & 809 & 1.96 (6) \\
    10 & \textbf{Distinct CVEs} & \multicolumn{2}{c}{\textbf{851}} & \multicolumn{2}{c}{\textbf{78}} \\
    \bottomrule
    \end{tabular}}
  \end{center}
\end{table}

Among the 5,068,635 blobs with newer versions, we find that $17.31\%$
and $14.36\%$ of the blobs were updated by a fix commit.
If we extrapolate the results, we see that 
in total, 101M blobs in the current \textit{full} Stack v2 database
(representing $17.30\%$ of all blobs in it) can be updated
to newer versions and $17.31\%$ of these new versions are bug fixes.
For the \textit{smol} dataset, we have 9M
(representing $10.18\%$ of all blobs in it) that can be updated
to newer versions and $14.36\%$ of those are bug fixes. While
deduplication reduced the proportion of buggy samples, millions of
them still remain and can be easily fixed.   

Finally, we checked how many code sample have fixes to known
vulnerabilities. To do that we searched for the regular expression
representing CVE ``cve-[0-9]+-[0-9]+'' and found that $2.36\%$ and
$1.88\%$ of the fixes in our sample relate to a known CVE.

Due to the important nature of known vulnerabilities, we further 
analyzed the complete \textit{smol} dataset---that is supposed to be most 
reliable version of the Stack v2---to find blobs that have
a newer version with fixes to known CVEs.
The results are shown in Table~\ref{t:cve}. We found that 19,944 blobs in
the \textit{smol} dataset have newer versions
that fixing a known CVE. These samples
were changed by 11,907 commits that mentioned 6,947 distinct CVEs in
their commit message. 

\begin{table}[t]
  \begin{center}
    \caption{CVE counts in complete smol dataset}
    \label{t:cve}
    \begin{tabular}{l|rrr}
    \toprule
    & CVE commits & CVE blobs & Distinct CVEs \\
    \midrule
    Count & 11,907 & 19,944 & 6,947 \\
    \bottomrule
    \end{tabular}
  \end{center}
\end{table}

In summary, despite careful curation and employment of sophisticated
heuristics, even the clean version of the Stack v2 dataset contains
millions of unfixed versions of the code, including thousands of
unfixed vulnerabilities that supports our first hypothesis (\textbf{H1}).

\begin{figure}[ht]
\begin{tcolorbox}[colback=gray!5!white, colframe=gray!70!black, title=Key Findings 1]
\begin{enumerate}[wide, labelwidth=1.5em, labelindent=0pt, leftmargin=*]
    \item 17.30\% and 10.18\% of blobs in the \textit{full} and \textit{smol} datastes, respectively, have newer versions, out of which 17.31\% and 14.36\% are bug fixes.
    \item 61.78\% and 55.32\% of blobs are the first version created, out of which 94.51\% 
    and 95.39\% have no newer versions, meaning they were created but never modified, 
    suggesting low quality.
    \item There are 19,944 blobs in the clean and deduplicated version of the 
    Stack v2 (\textit{smol}) that have a newer version were a known security vulnerability is being fixed.
    \item In total, 6,947 known CVEs has been found in the \textit{smol} dataset.
\end{enumerate}
\end{tcolorbox}
\end{figure}

\subsection{Potential Noncompliance}

The Stack v2 dataset consists of code that is either licensed under 
permissive terms or lacks a specified license. To address potential 
licensing concerns, the Stack v2 allows authors to opt out of inclusion 
in the dataset. It is important to note that code without a license 
is distinct from unlicensed code. From a copyright perspective, 
code without a license defaults to ``all rights 
reserved''~\cite{us_copyright_office_circular1},
which raises significant concerns about the inclusion of such 
code in this dataset.

As detailed in Section~\ref{s:noncompliance}, we analyzed blobs 
within the dataset that were reused across multiple OSS projects, 
as identified through WoC~\cite{jahanshahi2024dataset}. For each blob, 
we determined its originating project---the project with the earliest 
commit timestamp containing that blob---and cross-referenced it with 
the corresponding project in the Stack dataset. The results are
shown in Table~\ref{t:reuse}.

\begin{table}[t]
    \centering
    \caption{Reused blobs and their origin}
    \begin{tabular}{ll|rrrr}
         \toprule
          & & \multicolumn{2}{c}{\textbf{full}} & \multicolumn{2}{c}{\textbf{smol}} \\
          & & count & \% (row) & count & \% (row) \\
         \midrule
         1 & \textbf{Total} & \multicolumn{2}{c}{\textbf{582,933,549}} & \multicolumn{2}{c}{\textbf{87,175,702}} \\
         2 & Reused & 90,303,809 & 15.49 (1) & 9,848,987 & 11.30 (1) \\
         \midrule
         3 & Same & 29,432,636 & 32.59 (2) & 3,764,702 & 38.22 (2) \\
         4 & Different & 60,871,173 & 67.41 (2) & 6,084,285 & 61.78 (2) \\
         \bottomrule
    \end{tabular}
    \label{t:reuse}
\end{table}

The results indicate that $15.49\%$ and $11.30\%$ of blobs were reused
at least once. Furthermore, in $67.42\%$ and $61.78\%$ of instances, 
the originating projects identified by the Stack dataset differ 
from those identified by WoC. This highlights the inherent complexity 
of tracing the origins of code reused through copy-and-paste. 
WoC's ability to perform such identification stems from its 
comprehensive coverage of nearly all open-source projects 
and their version histories.

Since cases with misidentified origins present a potential risk of 
license noncompliance, we conducted a further investigation into the 
blobs with differing identified origins. The detailed results of 
this analysis are presented in Table~\ref{t:license}.

\begin{table}[t]
    \centering
    \caption{Reused blobs with different origins and their licenses}
    \resizebox{0.99\linewidth}{!}{
    \begin{tabular}{lll|rrrr}
         \toprule
         & & & \multicolumn{2}{c}{\textbf{full}} & \multicolumn{2}{c}{\textbf{smol}} \\
         & \textbf{Stack v2} & \textbf{WoC} & count & \% (row) & count & \% (row) \\
         \midrule
         1 & \multicolumn{2}{c|}{\textbf{Different Origin}} & \multicolumn{2}{c}{\textbf{60,871,173}} & \multicolumn{2}{c}{\textbf{6,084,285}} \\
         \midrule
         2 & \multicolumn{2}{c|}{\textbf{Same License}} & \textbf{38,410,728} & \textbf{63.10 (1)} & \textbf{4,418,289} & \textbf{72.62 (1)} \\
         3 & no license & no license & 26,604,621 & 69.26 (2) & 3,269,149 & 73.99 (2) \\
         4 & permissive & permissive & 11,806,107 & 30.74 (2) & 1,149,140 &  26.01 (2) \\
         \midrule
         5 & \multicolumn{2}{c|}{\textbf{Different License}} & \textbf{22,460,445} & \textbf{36.90 (1)} & \textbf{1,665,996} & \textbf{27.38 (1)} \\
         6 & permissive & no license & \textcolor{goldenrod}{10,257,891} & 45.67 (5) & \textcolor{goldenrod}{721,920} & 43.33 (5) \\
         7 & no license & permissive & \textcolor{forestgreen}{9,309,959} & 41.45 (5) & \textcolor{forestgreen}{658,085} & 39.50 (5) \\
         8 & no license & restrictive & \textcolor{red}{1,868,500} & 8.32 (5) & \textcolor{red}{193,358} & 11.61 (5) \\
         9 & permissive & restrictive & \textcolor{red}{1,024,095} & 4.56 (5) & \textcolor{red}{92,633} & 5.56 (5) \\
         \bottomrule
    \end{tabular}}
    \label{t:license}
\end{table}

The results reveal that $36.90\%$ and $27.38\%$ of the blobs with 
misidentified origins have licenses that differ from those 
identified in the Stack dataset. These discrepancies fall into four 
distinct categories. In the first case, the Stack identifies the 
license as permissive, while WoC identifies no license. 
In the second, the Stack identifies no license, but WoC identifies 
a permissive license. The third case involves the Stack identifying 
no license, while WoC identifies a restrictive license. Finally, 
in the fourth case, the Stack identifies a permissive license, 
but WoC identifies a restrictive license. Among these, the second 
scenario does not pose a compliance risk and may even be advantageous, 
given the problematic nature of reusing code without a license, as 
previously discussed. However, the first scenario still raises some 
concerns. The third and fourth scenarios are particularly concerning 
as they indicate a high risk of license noncompliance due to the 
blobs originating from projects with restrictive licenses.

In summary, our analysis reveals that even the smaller version of the 
Stack dataset contains hundreds of thousands of blobs originating from 
projects with restrictive licenses, raising significant legal compliance 
concerns for any LLM trained on this dataset. These findings provide 
strong support for our second hypothesis (\textbf{H2}).

\begin{figure}[ht]
\begin{tcolorbox}[colback=gray!5!white, colframe=gray!70!black, title=Key Findings 2]
\begin{enumerate}[wide, labelwidth=1.5em, labelindent=0pt, leftmargin=*]
    \item 15.49\% and 11.30\% of blobs in the \textit{full} and \textit{smol} datasets, 
    respectively, have been reused at least once. Among these, 64.41\% and 61.78\% have 
    origins that were misidentified.
    \item 36.90\% and 27.38\% of blobs with misidentified origins have licenses that 
    differ from those identified in the dataset.
    \item 12.88\% and 17.17\% of blobs with differing licenses are subject to a 
    restrictive license, presenting a significant risk of noncompliance.
\end{enumerate}
\end{tcolorbox}
\end{figure}

\section{Limitations}

\subsection{Internal Validity}

\subsubsection{Impact of Buggy Code Removal on Model Outputs}
Eliminating all buggy code from pre-training or fine-tuning datasets 
does not guarantee that the resulting LLM will generate bug-free code. 
However, it is reasonable to assume that some generated code may 
replicate buggy patterns observed in the training data. Therefore, 
removing bugs from the training data, especially through a low-cost 
approach like ours, is a sensible step toward improving the model's 
output quality.

\subsubsection{WoC Dataset Coverage}
Some code may originate outside public version control systems or 
may simply not be included in WoC's collection. However, as demonstrated 
with the Stack v2 dataset, only $2.5\%$ of blobs could not be linked 
to commits already present in WoC, indicating that this is a relatively 
minor issue.

\subsubsection{Blob Updates and Quality}
While updating blobs to newer versions eliminates known bugs, 
it can occasionally introduce new and unknown bugs. However, 
in most projects, only a small proportion of bug fixes result 
in new issues or fail to address the intended bugs. Consequently, 
applying fixes generally enhances the overall quality of 
the training data.

\subsubsection{Rebasing and Metadata Loss}
Our approach relies on git SHA-1 hashes to track blobs, 
which ensures that content-based identification is robust to rebasing.
However, rebasing may obscure certain metadata, such as precise commit 
lineage, which could limit the ability to fully trace the historical 
context of some blobs.

\subsubsection{Commit Keyword Usage for Fix Identification}
Not all commits containing the keywords we used represent bug fixes, 
nor do all bug fixes include these keywords in their commit messages.
Despite this, applying all changes, not just those identified as fixes,
is likely necessary. These keywords and similar ones have been widely
used in prior research to identify changes related to bug fixes. 
In our validation of 20 randomly selected commits, only three 
($15\%$) were found not to be clearly bug fixes.

\subsubsection{Reliability of CVE Detection}
Our method successfully identified thousands of CVEs in the Stack v2 
dataset, leveraging commit messages as a primary indicator. However, 
this approach relies on the presence of explicit references to CVEs in 
commit messages, which may not comprehensively capture all vulnerabilities.
For instance, CVEs that were not documented in commit messages or 
introduced through transitive dependencies might be missed. Future 
work could address this limitation by conducting a manual review of a 
representative sample or validating the method against additional 
datasets to evaluate recall more comprehensively.

\subsection{Construct Validity}

\subsubsection{Impact of Dataset Vulnerabilities on Model Outputs}
This study assumes that vulnerabilities and flaws in training 
datasets may influence the quality and security of model outputs.
While this assumption aligns with logical inference and prior 
research on LLM behavior, direct empirical validation of this 
relationship is currently lacking and represents an important 
avenue for future research.

\subsubsection{Never-Modified Code Assumption}
While we suggest that never-modified code may indicate
low use or untested quality, this is based on logical inference rather 
than direct empirical evidence. Future studies are needed to validate 
whether unmodified code consistently correlates with lower 
reliability or usability in practice.

\subsubsection{Blob Origin Identification}
Identifying the origin of a blob is not always possible, particularly 
for blobs that did not originate in open-source projects. Accurate 
identification requires comprehensive access to all project data. 
However, the extensive coverage provided by WoC significantly 
reduces this risk.

\subsubsection{License Applicability Assumption}
The licensing assumption for a blob is based on the identified license
of the project from which it originated. However, not all files 
within a project necessarily fall under the project's overarching 
license, as some files may have distinct individual licenses.

\subsection{External Validity}

\subsubsection{New Bugs and Iterative Updates}
Even if all known bugs are addressed at time $t$, new bugs will 
inevitably be discovered at time $t+1$. Therefore, regular 
updates are necessary. Fortunately, the approach outlined 
here can be automated, allowing it to be efficiently applied 
to each new version of the WoC dataset.

\subsubsection{Updating to Latest Versions}
The updated version of a blob may not always represent the 
latest available version. As a result, the process may need to 
be repeated iteratively until the most recent fix is applied.
The median timestamp of the commits updating blobs was June 2020, 
indicating that these updates were available well before the 
creation of the Stack v2 dataset in 2024.

\section{Future Work}

A promising direction for future work is the development of 
automated curation tools specifically designed to enhance the 
quality of datasets used for pre-training large language models (LLMs) 
for code, such as Stack v2. Building on the cost-efficient 
approach introduced in this paper, these tools could automatically 
identify and apply patches for known fixes or vulnerabilities, 
ensuring that the datasets include secure and reliable code. 
They could also locate and update blobs to their latest versions, 
minimizing the inclusion of outdated or buggy code. 
Furthermore, the tools could enhance license compliance by 
automatically detecting and removing code with non-permissive
licenses, ensuring that only code with appropriate 
licensing is included in the dataset. 
The feasibility of such automation is demonstrated by the 
scalability and efficiency of our approach in handling large-scale datasets. 
By automating these tasks, the proposed tools would streamline 
the iterative updates required for maintaining high-quality training 
data, ensuring practicality and cost-effectiveness in preparing 
datasets for LLM pre-training.

\section{Conclusions}

Processes to ensure provenance, security, and compliance in SSCs are
essential. This project sets the stage for future work on the
curating LLM training data and provide several insights and
interventions that can improve on the current state of the art.

Several notable observations emerge from our analysis. First, 
the largest open-source training dataset, Stack v2, contains only 
a small fraction of all publicly available source code versions. 
These datasets could be significantly enhanced by incorporating 
intelligently selected data from comprehensive sources like WoC. 
Second, between 10\% and 20\% of the versions have updates, even 
though the WoC dataset version V3 is contemporaneous with Stack v2.
Third, a substantial portion of the training data includes files
with known bug fixes. While newer versions may incorporate updated 
APIs or additional features, applying these bug fixes is crucial
to prevent LLMs from being trained on buggy code. 
Fourth, such fixes can be leveraged to train or align LLMs that
specialize in generating changes or fixes. 
Fifth, training datasets should prioritize heavily or moderately
modified code, which often has fewer bugs, rather than relying 
heavily on pristine, first-version code that dominates many 
existing datasets. Finally, misidentified code origins have
resulted in non-permissive code being included in these datasets, 
raising compliance concerns.

Beyond improving the curation practices for LLM training data, 
this work also introduces the concept of the LLM supply chain, 
highlighting its similarities to and differences from traditional 
software supply chains.

While our primary focus has been on data curation for code LLMs, 
the insights generalize to any scenario involving version-controlled data.

\balance
\small
\bibliographystyle{IEEEtranN}
\setlength{\bibsep}{0pt plus 0.3ex}  % Adjust bibliography spacing to mimic IEEE format
\bibliography{ref}

\end{document}